\documentclass[runningheads,a4paper]{llncs}

\usepackage{amssymb}
\usepackage{graphicx}
\usepackage{multirow}

\usepackage{url}
\urldef{\mailsa}\path|ghaemmaghami.pouya@gmail.com|

\begin{document}

\mainmatter  

\title{Reconstruction of the External Stimuli from Brain Signals}

\titlerunning{Reconstruction of the External Stimuli from Brain Signals}

%
%
\author{Pouya Ghaemmaghami\inst{1}}

\authorrunning{Ghaemmaghami et al.}

\institute{\mailsa\\
}


%
%

\toctitle{Towards Mind Reading}
\tocauthor{Authors' Instructions}
\maketitle

\begin{abstract}
	Despite the rapid advances in Brain-computer Interfacing (BCI) and continuous effort to improve the accuracy of brain decoding systems, the urge for the systems to reconstruct the experiences of the users has been widely acknowledged. This urge has been investigated by some researchers during the past years in terms of reconstruction of the naturalistic images, abstract images, video and audio. In this study, we try to tackle this issue by regressing the stimuli spectrogram using the spectrogram analysis of the brain signals. The results of our regression-based method suggest the feasibility of such reconstructions using the neuroimaging techniques that are appropriate for out-of-lab scenarios.
\end{abstract}

	\section{Introduction}
	\label{sec:rec_intro}
	Extracting stimulus-related information from the brain activity using machine learning algorithms is known as brain decoding which has been used widely in neuroscience community in the past decade. Prior works on brain decoding have mostly focused on the classification of the stimuli into a set of pre-defined categories \cite{haxby2001distributed,cox2003functional,mitchell2004learning,wang2009brain,carlson2011high,ghaemmaghami2015movie,ghaemmaghami2016music}. A typical classification pipeline includes the following steps: First, different categories of stimuli are presented to the participant of the experiment, while his/her concurrent brain activity is recorded. Then a machine learning algorithm is trained on the subset of the samples in order to learn the mapping function between the brain activity pattern and the stimulus category from a training data set. If the algorithm, can predict the target stimulus category of the remaining subset (test-set) better than the chance level, we can hypothesize that the stimulus-related information is encoded in the brain data.

\begin{figure}[t!]
\centering
\includegraphics[width=\linewidth]{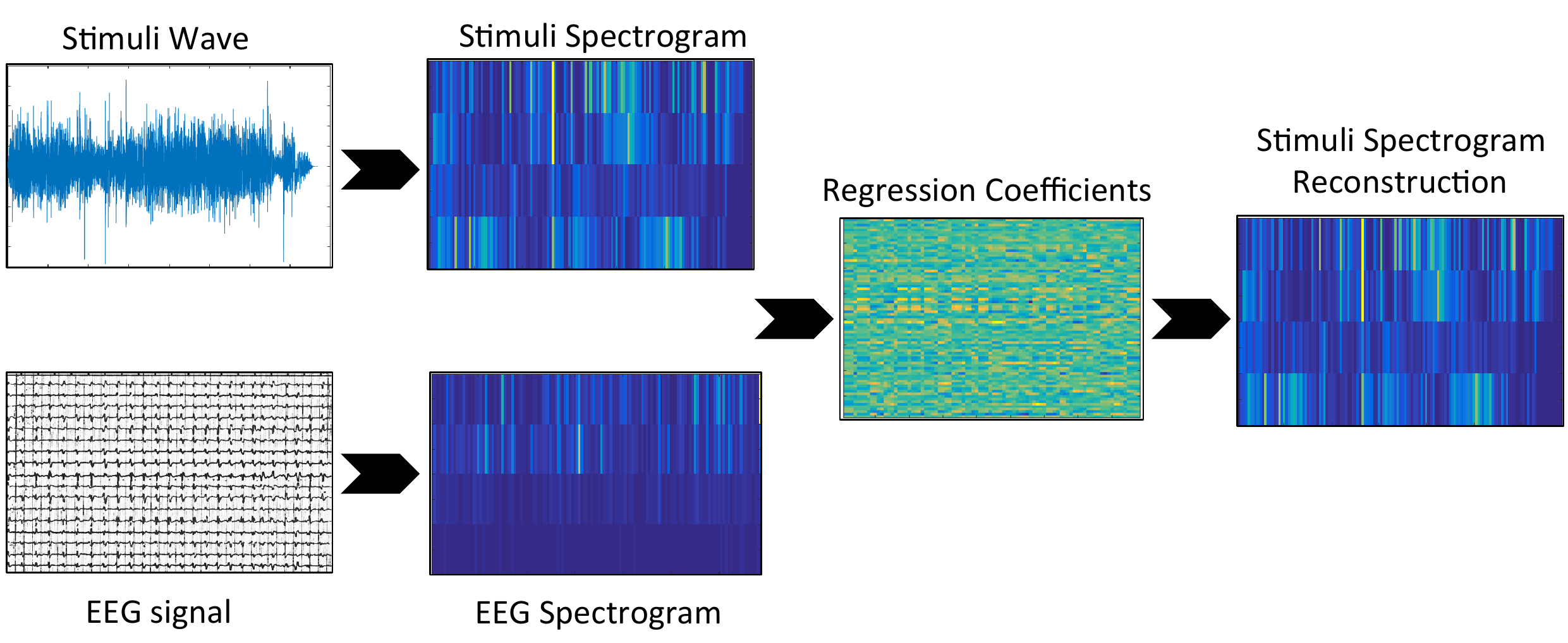}
\caption{\textbf{Stimuli reconstruction pipeline:} In the first step of the analysis the spectrogram of the EEG signal and audio wave (after generic preprocessing steps) is computed. Then, in the training phase, the EEG features are regressed (Ridge Regression) onto the audio spectrogram to find the proper mapping function. After that, the resulting weight matrix is used on the test EEG data to predict the spectrogram of the audio wave.}
\label{fig:rec_framework}
\end{figure}

However, such classification approaches are not able to capture all aspects of the stimulus, since they are constrained by the stimulus category. Instead, a more challenging scenario can be the stimulus' reconstruction that aims at recreating the stimulus using the recorded brain activity, free from the constraint of the categories \cite{miyawaki2008visual}. Such scenario has been investigated by some researchers during the past years in terms of reconstruction of the naturalistic images \cite{kay2008identifying,naselaris2009bayesian}, abstract images \cite{kuo2014decoding}, video \cite{nishimoto2011reconstructing} and audio \cite{pasley2012reconstructing}. However, almost all of these studies are conducted using either invasive methods or bulky expensive non-invasive methods such as fMRI (Functional magnetic resonance imaging) or MEG (Magnetoencephalography) which are not appropriate for out-of-lab scenarios. In this study, instead, we try to tackle this issue using EEG (Electroencephalography) that is the most commonly used signal acquisition technique in BCI. In particular, we aim at regressing the stimuli spectrogram using the spectrogram analysis of the EEG signal. Figure \ref{fig:rec_framework} illustrates the overall framework used in our study. To our knowledge, we are one of the first showing the possibility of reconstructing stimuli spectrogram using EEG signals.

The remainder of this study is organized as follows: Section \ref{sec:rec_litreature} reviews related literature on this topic. Section \ref{sec:rec_materials} explains the employed dataset and the data analysis methods used for data compilation. The results of such analysis are presented and discussed in Section \ref{sec:rec_results}. Finally Section \ref{sec:rec_conc} concludes this paper with the key observations and some possible future directions. \\
	
	\section{Literature Review}
	\label{sec:rec_litreature}
	Reconstruction of someone's experiences from his/her brain activity patterns can be considered as the ultimate goal of ``Mind Reading". During the past recent years, many researchers tried to tackle this task in many different ways using different neuroimaging modalities and different methodologies. Here, we briefly review some of the major works on this topic.

\subsection{Reconstructing image-based stimuli}
Reconstructing image-based stimuli has received more attention compared to the other types of stimuli. In a very typical way, brain responses are used for building an encoding model for each specific brain regions. Such model can describe the dependency between a brain region and a particular set of stimuli features. Once these models are learned, they can be employed to predict the brain activity patterns of the new stimuli. Such prediction can be validated by correlating the predicted brain response with the original brain response. This procedure was employed in numerous studies in order to reconstruct images \cite{kay2008identifying,naselaris2009bayesian}, movies \cite{nishimoto2011reconstructing}, remembered scenes \cite{stansbury2013natural,naselaris2015voxel} and handwritten characters \cite{schoenmakers2013linear}. However, in order to improve the quality of recosntruction, the authors generally  used a database of priors \cite{kay2008identifying,naselaris2009bayesian,nishimoto2011reconstructing}, hence as authors in \cite{nishimoto2011reconstructing,schoenmakers2013linear} conclude, the exact reconstruction is not possible and the quality of such reconstructions are highly dependent to the number and the quality of the priors. In other words, changing the prior, can yield to a completely different reconstruction.

Unlike these studies, Thirion et al. \cite{Thirion20061104}, Miyawaki et al. \cite{miyawaki2008visual} and Kuo et al. \cite{kuo2014decoding} tried to reconstruct visual stimuli from brain activity, without any kind of image priors. This is based on the assumption that there exists a function that maps the visual stimuli with the corresponding brain responses elicited from such stimuli. By inverting this function, the stimuli can be reconstructed given the brain activity patterns. However, authors just used very simple-abstract images with limited shapes. Even in such case, the correlation coefficient between the reconstruction and the original stimulus was not significant for all stimulus and all subjects. 

\subsection{Reconstructing audio-based stimuli}
Reconstructing audio-based stimuli has received less attention compared to the visual types of stimuli (i.e. images and videos). However, recently, researchers began to explore this type of stimuli as well. In \cite{pasley2012reconstructing,martin2014decoding}, authors tried to reconstruct the spectro-temporal auditory features from neural responses using an invasive neuroimaging technique (ECoG). They obtained significant correlation, between the reconstructed spectrogram and the original wave spectrogram.  

Strum et al., \cite{sturm2015multi}, tried to reconstruct the musical stimuli power-slope using a non-invasive approach (EEG). Using a Ridge Linear Regression approach, they were able to regress the temporally embedded EEG features into power-slope of the musical stimuli. However in most of the cases the correlation coefficient is very low and insignificant.

In an interesting and recent work by Huth et al. \cite{huth2016natural}, authors tried to build a semantic model for each voxel by regressing the brain activity patterns of each voxel with the semantic features obtained from more than two hours of narration of the stories. Such analysis revealed high correlation in some voxels, but the average performance (in between-subject analysis) is rather weak (less than 0.08). Although authors did not investigate the possibility of reconstruction of the new stimuli (narration of the story) from the brain responses, however, using such semantic voxel models might enable them to do so in further experiments. 

\subsection{Spotting the gap}
Our examination of the related literature reveals that reconstructing the experiences of the users are recently capturing attention across many communities and will be a hot topic with the continuous advances in signal acquisition techniques and machine learning algorithms. However, these studies are conducted using the neuroimaging techniques that are not appropriate for out-of-lab scenarios (except the work of \cite{sturm2015multi} which is explained above and the works of \cite{kim2014reconstruction,korik2015e3d} on hand movement reconstrcution). Since EEG is probably the most common signal acquisition technique in BCI, hence, it would be interesting to investigate whether the EEG responses to the stimuli can be utilized to obtain information regarding the stimulus or not. In light of this, in the present study, we propose an approach to regress the spectrogram of the stimulus using spectrogram analysis of the EEG signals in order to complement a possible link between stimulus structure and brain signals. \\


	\section{Materials and Methods}
	\label{sec:rec_materials}
	In this section, we describe the employed dataset, feature extraction
scheme, and data analysis procedure. 

\subsection{Dataset} In our experiments, we used DEAP dataset \cite{koelstra2012deap} that contains the electroencephalographic (EEG) data of 32 participants who watched 40 music video clips. The music video clips were projected onto a screen at a screen refresh rate of 60 Hz. The electroencephalographic data were recorded at a sampling rate of 512 Hz using a 32 channel device.

\subsection{Data Analysis}	

\subsubsection{EEG Signal Processing:} 
In this study, the pre-processed EEG data in \cite{koelstra2012deap} is employed and processed further. The whole processing steps are as follows:
\begin{enumerate}
  \item Down-sampling the EEG signal to 128 Hz. 
  \item EOG artifacts removal.
  \item Bandpass frequency filtering (4 - 45 Hz).
  \item Estimating the spectral power of each channel of the EEG trials with a window size of 64 samples (500 milliseconds) with 8 samples overlap between windows.
  \item Calculating the stimuli spectrogram by averaging the signal power over four major frequency bands: theta (3:7 Hz), alpha (8:15 Hz), beta (16:31 Hz) and gamma (32:45 Hz).
\end{enumerate}

The output of this procedure for each trial is a matrix with the following dimensions: 32 (number of the EEG sensors) $\times$ 4 (major frequency bands) $\times$ 137 (number of segmented temporal windows). \\



\noindent\textbf{Sensor Selection:} Not all brain regions involve in processing the auditory information in humans. The auditory cortex that is located bilaterally at the top of the temporal lobes, involves in perception and understanding of voices \cite{belin2000voice,zatorre2002structure}. In light of this, in this study, we used the sensors located in the temporal area of the brain (T7 and T8). In order to increase the signal to noise ratio, for each subject and on each clip, we averaged the time-frequency outputs of these two sensors. 


\subsubsection{Stimuli Processing:} For each stimulus (music video clip), the signal power is determined as follows: 

\begin{enumerate}
  \item Downsampling the audio signal to the sampling frequency of the EEG signal (128 Hz).
  \item Segmenting the audio signal into the 12.5 \% overlapping time frames of 500 milliseconds width.
  \item Estimating the average signal power for each window for every frequency.
  \item Calculating the stimuli spectrogram by averaging the signal power over four frequency bands: theta (3:7 Hz), alpha (8:15 Hz), beta (16:31 Hz) and gamma (32:45 Hz). The output of this procedure is a 3-dimensional matrix with the following dimensions: 40 (number of the music video clips) $\times$ 4 (major frequency bands) $\times$ 137 (number of segmented temporal windows).
\end{enumerate}

\subsection{Regression Analysis}\label{regression_procedure}
We adopted Linear Ridge Regression model (\ref{eq:1}) under leave-one-sample-out cross-validation schema in order to minimize the distance between the spectrogram of the EEG signals and the spectrogram of the audio signal. Thus, for a given stimulus on a particular subject a mapping function ($\beta$) is learned based on the EEG/Audio spectrogram of 39 stimulus in the training set ($X_{train}$). Then, the mapping function is applied to the EEG spectrogram of the remaining stimulus in the test set ($X_{test}$) in order to predict the spectrogram of the audio stimulus ($Y$). Such procedure is repeated 40 times so that each stimulus for each subject is once used in the test set. \\

\begin{equation} \label{eq:1}
\hat{\beta}^{ridge} = argmin\left \| Y-\beta X  \right \|_{2}^{2} + \lambda \left \| \beta  \right \|_{2}^{2}
\end{equation}

	\section{Results}
	\label{sec:rec_results}
The prediction of the spectrogram/slope of the audio stimulus from EEG signals is served as the basis for examining the relation between the stimulus and the reconstructed version of the stimulus. In order to do so, we performed the following two experiments:

\subsection{Experiment 1: Reconstructing Spectrogram}
In the first experiment, we aimed at reconstructing the spectrogram of the audio stimulus from the EEG signals. On this, the audio wave is smoothed using a Gaussian filter ($\sigma=2$). Then, as explained in the previous section, we employed a ridge regression model for such task. Under leave-one out cross validation scheme, the model was trained using the 39 stimuli, and was tested on the remaining stimulus. This has been done 40 times so that each stimulus was tested one time. The ridge parameter ($\lambda$) in the ridge regression model was tuned using a 5-fold cross validating in the ``training set" by employing a range of values. Once, the audio signal (regarding the stimulus in the test set) is reconstructed using the EEG features, we calculate the Pearson correlation between the spectrogram of the stimulus (in the test set) and the reconstructed spectrogram of the same stimulus using the EEG signals. This procedure, finds 40 correlation coefficient (r-values and their corresponding p-values) for each stimulus of each subject. To compare the results on a subject level basis, the correlation coefficients are averaged for each subject. Accordingly, for each subject, the obtained p-values are fused over all clips using the Fisher's method \cite{fisher1956statistical,bailey1998combining}. The results of such analysis are demonstrated in Table~\ref{tab:correlation_analysis}.

\subsection{Experiment 2: Reconstructing Slope}
In the second experiment, following \cite{sturm2015multi}, we aimed at reconstructing the slope of the audio stimulus from the EEG signals. On this, the audio signal power is averaged over all frequency bands (theta, alpha, beta and gamma). Subsequently, the first derivative is taken and the resulting signal is smoothed using a Gaussian filter ($\sigma=2$). Then, the same Linear Ridge Regression model under leave-one-sample-out cross-validation schema is adopted in order to predict the audio slope given the EEG signals spectrogram. The rest of the correlation analysis is the same the first experiment. Table~\ref{tab:correlation_analysis} demonstrates the results of such analysis. \\

\begin{table}[t!]
\centering
\fontsize{12}{12}\selectfont
\caption{\label{tab:correlation_analysis} Correlation analysis.}
\begin{tabular}{|c|c|c|c|c|}
\hline
\multirow{2}{*}{\textbf{Subjects}} & \multicolumn{2}{c|}{\textbf{Audio-Spectrogram}} & \multicolumn{2}{c|}{\textbf{Audio-Slope}}\\ \cline{2-5} 
& \textbf{r-value} & \textbf{p-value} & \textbf{r-value} & \textbf{p-value}\\ \hline
Sub 1 & 0.163 & $<$ 0.001 & 0.806 & $<$ 0.001\\
Sub 2 & 0.106 & $<$ 0.001 & 0.550 & $<$ 0.001\\
Sub 3 & 0.081 & $<$ 0.001 & 0.604 & $<$ 0.001\\
Sub 4 & 0.016 & $<$ 0.001 & 0.504 & $<$ 0.001\\
Sub 5 & 0.117 & $<$ 0.001 & 0.590 & $<$ 0.001\\
Sub 6 & 0.101 & $<$ 0.001 & 0.777 & $<$ 0.001\\
Sub 7 & 0.158 & $<$ 0.001 & 0.814 & $<$ 0.001\\
Sub 8 & 0.121 & $<$ 0.001 & 0.609 & $<$ 0.001\\
Sub 9 & 0.099 & $<$ 0.001 & 0.666 & $<$ 0.001\\
Sub 10 & 0.097 & $<$ 0.001 & 0.714 & $<$ 0.001\\
Sub 11 & 0.079 & $<$ 0.001 & 0.483 & $<$ 0.001\\
Sub 12 & 0.099 & $<$ 0.001 & 0.745 & $<$ 0.001\\
Sub 13 & 0.094 & $<$ 0.001 & 0.554 & $<$ 0.001\\
Sub 14 & 0.085 & $<$ 0.001 & 0.548 & $<$ 0.001\\
Sub 15 & 0.104 & $<$ 0.001 & 0.620 & $<$ 0.001\\
Sub 16 & 0.154 & $<$ 0.001 & 0.661 & $<$ 0.001\\
Sub 17 & 0.070 & $<$ 0.001 & 0.517 & $<$ 0.001\\
Sub 18 & 0.112 & $<$ 0.001 & 0.776 & $<$ 0.001\\
Sub 19 & 0.109 & $<$ 0.001 & 0.554 & $<$ 0.001\\
Sub 20 & 0.163 & $<$ 0.001 & 0.764 & $<$ 0.001\\
Sub 21 & 0.136 & $<$ 0.001 & 0.466 & $<$ 0.001\\
Sub 22 & 0.040 & $<$ 0.001 & 0.408 & $<$ 0.001\\
Sub 23 & 0.114 & $<$ 0.001 & 0.790 & $<$ 0.001\\
Sub 24 & 0.107 & $<$ 0.001 & 0.706 & $<$ 0.001\\
Sub 25 & 0.083 & $<$ 0.001 & 0.623 & $<$ 0.001\\
Sub 26 & 0.111 & $<$ 0.001 & 0.550 & $<$ 0.001\\
Sub 27 & 0.114 & $<$ 0.001 & 0.733 & $<$ 0.001\\
Sub 28 & 0.097 & $<$ 0.001 & 0.747 & $<$ 0.001\\
Sub 29 & 0.069 & $<$ 0.001 & 0.522 & $<$ 0.001\\
Sub 30 & 0.102 & $<$ 0.001 & 0.629 & $<$ 0.001\\
Sub 31 & 0.092 & $<$ 0.001 & 0.717 & $<$ 0.001\\
Sub 32 & 0.110 & $<$ 0.001 & 0.639 & $<$ 0.001\\
\hline \hline
\textbf{Average} & 0.103 & $<$ 0.001 & 0.637 & $<$ 0.001\\  \hline
\end{tabular}
\end{table}

	\section{Conclusion}
	\label{sec:rec_conc}
	In this study, we presented an approach regarding reconstruction of the audio stimulus spectrogram directly from the EEG brain signals. The results of our regression-based method demonstrate the feasibility of the reconstruction of the spectrogram of the audio stimulus directly from the EEG signals at the single-subject level. The obtained correlation coefficients are significant (p-value $<$ 0.001). The presented results are just a proof-of-concept that multivariate analysis of the brain signals may extract complex stimuli related information from brain. To our knowledge, this study is one of the first efforts that employs EEG signals for such tasks. Nevertheless, several issues call for further explorations. First and foremost, the obtained correlation coefficient is still rather weak (although it is significant and it is consistent with the correlation value reported in the other literatures \cite{martin2014decoding,sturm2015multi,huth2016natural}). On this, other regression methods can be explored. One of the promising technique is applying canonical correlation in order to find a subspace to maximize the correlation between two representations. Another useful approach is to employ cross-modal domain adaptation approaches in order to transfer knowledge from multimedia domain (well-performing domain) to the brain domain (poor-performing domain). Recently, it has been shown that such approaches leads to improved results in certain classification tasks in the brain domain \cite{ghaemmaghami2016brain,ghaemmaghami2017adaptation}). Secondly, the variance of the correlation coefficient between stimuli within (and between) subjects has not been explained yet and should be explored in future experiments. Thirdly, other characteristics of the music stimulus (rather the signal spectrogram) need to be examined in the subsequent studies. In addition, it would be interesting to probe whether or not such stimuli can cause significant correlation with physiological responses, such as heart rate. \\


\bibliography{rec}
\bibliographystyle{splncs03}

\end{document}